# Formation and manipulation of diatomic rotors at the symmetry-breaking surfaces of kagome superconductors

Zihao Huang[1,2,#], Xianghe Han[1,2,#], Zhen Zhao[1,2], Haitao Yang[1,2], Hui Chen[1,2,*] and Hong-Jun Gao[1,2,*]

[1] Beijing National Center for Condensed Matter Physics and Institute of Physics, Chinese Academy of Sciences, Beijing 100190, PR China

[2] School of Physical Sciences, University of Chinese Academy of Sciences, Beijing 100190, PR China

# ABSTRACT

Artificial molecular rotors and motors hold great promise for functional nanomachines, but constructing diatomic rotors, crucial for these machines, is challenging due to surface constraints and limited chemical design. Here we report the construction of diatomic Cr-Cs and Fe-Cs rotors where a Cr or Fe atom revolves around a Cs atom at the Sb surface of the newly-discovered kagome superconductor $CsV_3Sb_5$. The rotation rate is controlled by bias voltage between the rotor and scanning tunneling microscope (STM) tip. The spatial distribution of rates exhibits $C_2$ symmetry, might linked to the symmetry-breaking charge orders of $CsV_3Sb_5$. We have expanded rotor construction to include different transition metals (Cr, Fe, V) and alkali metals (Cs, K). Remarkably, designed configurations of rotors are achieved through STM manipulation. Rotor orbits and quantum states are precisely controlled by tunning inter-rotor distance. Our findings establish a novel platform for the atomically precise fabrication of atomic motors on symmetry-breaking quantum materials, paving the way for advanced nanoscale devices.

The significance of nanoscale machines in diverse biological processes has spurred advancements in drug delivery, microelectronics, and the manipulation of nanoscale materials [1-6]. A pinnacle pursuit in molecular-scale engineering is the creation of artificial molecular machines capable of controlled mechanical motion and intricate tasks [7-9]. In recent years, the synthesis of various artificial molecular machines, such as molecular gears [10], nanocars [11] and nanorobots [5], has opened up extensive possibilities for the future of nanotechnology. While significant attention has been devoted to the design, construction, manipulation, and assembly of molecular rotors and motors as foundational components for nanomachines, the focus has primarily been on larger structures [12-22]. Quantum diatomic rotors, comprising only two atoms, represent the smallest building blocks and offer versatile potential for constructing various nanostructures through precise design and assembly. Theoretically, the rotational behavior of an $H_2$ molecule on a solid surface provides an ideal model for the elemental rotor. However, practical implementation has encountered limitations, particularly concerning rotational behaviors restricted to ortho-para conversion and the edges of metal surfaces [23, 24]. The construction of diatomic rotors, a crucial and challenging task in the realm of molecular rotors and nanomachines, remains largely unexplored.

In recent years, selecting functional quantum materials as substrate provides an effective strategy for the construction of new type surface-mounted rotors at molecular level. Compared with conditional noble metal substrates [15], the quantum materials surface such as PdGa(111) [25], Bi(111) [26] and carbon nanotube [27] exhibit symmetry-breaking non-trivial surface states or specific surface confinements, leading to a unconventional molecule-substrate interactions and the movements of molecular rotors. Such strategy also inspires the searching for diatomic rotor. Very recently, Desvignes *et al.* reported a diatomic rotor in Fe doped $Bi_2Se_3$ where the atomically sized Fe defect embedded in the crystal structure [28].

Here we report, for the first time, the construction of a Cr-Cs/ Fe-Cs diatomic rotor in a newly-discovered kagome superconductor $CsV_3Sb_5$ [29, 30], which has been reported to exhibit rich surface effects such as symmetry breaking orders [31-36], carrier doping [37, 38] and polarity [39]. Using low-temperature scanning tunneling microscope (STM), we prove the diatomic nature of the rotors. The isolated alkali metal Cs atoms which originally stay at the Sb surface serves as the anchor, while the transition metal Cr/Fe atom that is evaporated onto the Sb surface traps into one Cs atom and rotate around the anchor. The anisotropic distribution of switching rates suggests the effects of symmetry breaking charge orders on the rotors. The rotation rate of the rotor changes as a function of bias voltage in the tunneling. We construct a series of diatomic rotors with various rotary properties by substituting the components of the diatomic rotors by other transition metals (such as Fe and V) or alkali metals (such as K). Remarkably, the rotation orbits and rates are manipulated in atomic precision by placing the rotors in different motif. Our works provide a new platform for exploring and manufacturing atomic rotors and nanodevices in novel quantum materials.

The schematic of construction process of Cr-Cs diatomic rotor is presented in Figure 1(a). Following cleavage of CsV$_3$Sb$_5$ at low temperature, Cs atoms exhibit diverse reconstructions on the as-cleaved surface [40]. As the externally-evaporated metal adatoms are difficult to be distinguished from the complex Cs reconstructions (Figure S1), we focus on the large-scale Sb terminated surface where the individual isolated Cs adatoms are randomly distributed (Figure S2). The Cr atoms are evaporated from the external metal sources onto the as-cleaved surface of CsV$_3$Sb$_5$ (details see Method part). Upon evaporating Cr atom onto the Sb surface, certain Cr atoms capture isolated Cs adatoms, forming the "rotors" in the tunneling process, as schematically shown in Figure 1(a).

In the large-sized STM topographic image, the rotors exhibit wheel-like protrusions (marked by the red dotted circles) which are surrounded by the randomly distributed and isolated Cs atoms (Figure 1(b)). The close-up STM image shows that the wheel-like protrusion consists of an inner spherical center and an outer annulus orbit (Figure 1(c)). The inner spherical center shows a stable topographic feature with a diameter of approximately 1.1 nm, while the annulus orbit shows unstable height signal in the constant-current-mode STM images (Figure S3).

To reveal the atomic configuration of the rotor, we apply a tip pulse to tear the rotor apart [13]. As a result, the rotor divides into two distinct atoms with different sizes (Figure 1(d)). As the larger atom shares a similar diameter and height with the isolated Cs adatoms at the Sb surface (Figure S3), we attribute the larger atom to the Cs atom, while assign the smaller atom to the Cr atom. The tip-induced splitting of the rotor into two separated atoms is reversible. After pushing a Cr and a Cs atom closer by an STM tip, the Cs and Cr atoms recombine, re-establishing the rotor configuration (Figure 1(c), (d)).

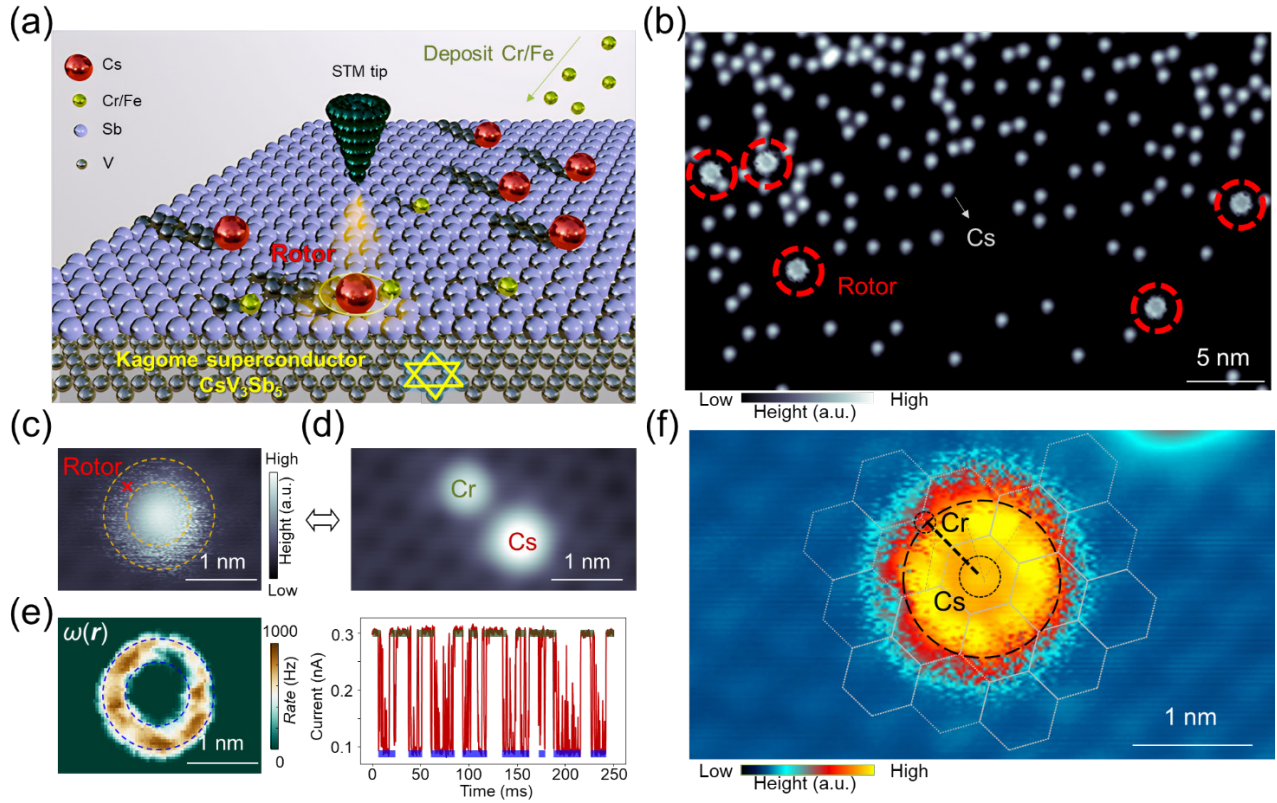

**Figure 1.** Construction of Cr-Cs diatomic rotors at the Sb surface of $CsV_3Sb_5$ with diluted Cs adatoms. (a), Schematics of the formation of diatomic rotors. The Cr/Fe atom (marked by the yellow ball) distribute as a single atom and trap into a Cs atom (marked by the red ball) to form a diatomic rotor on the Sb surface of kagome superconductor $CsV_3Sb_5$. (b), STM images showing the Sb surface of $CsV_3Sb_5$ crystal with dilute Cs atoms. The Cr-Cs diatomic rotors are highlighted by the red dotted circles ($V_s$=-500 mV, $I_t$=3 nA). (c) - (d), The tip-induced separation of Cr-Cs rotors into a Cr and a Cs atom. Before separation, the Cr atom revolve around Cs atom forming a Cr-Cs rotor with an unstable annulus (c). After separation, the morphography of Cs and Cr atom are clearly visible (d). $V_s$=-500 mV, $I_t$=3 nA. (e), Left: Rotation rate map $\omega(r)$ of (c), showing that Cr atom rotates around Cs in a circular orbit ($V$=-600 mV, $I$=0.5 nA). Right: The $I$-$t$ spectrum measured at the annulus position (marked by the red cross) of Cr-Cs rotor in (c), showing steps-like feature with several discrete values ($V$=-250 mV, $I$=0.9 nA). (f), Atomically-resolved STM topography of Cr-Cs diatomic rotor at the Sb surface of $CsV_3Sb_5$. The atomic model and Sb honeycomb lattice (white dotted hexagons) are overlaid in the image, showing that a Cr atom revolve around the Cs adatom ($V_s$=-500 mV, $I_t$=3 nA).

We then investigate the rotatory nature of the rotor by monitoring the evolution of tunneling current over time (*I-t* curve) at the annulus orbit part of the rotor while maintaining a constant tip height. The *I-t* spectrum obtained at the annulus orbit exhibits cyclic jump sequences between two major current levels (right panel of Figure 1(e)), indicating that the rotor undergoes rotational switching between two unequal configurations. To study the spatial distribution of rotating features, we conduct the *I-t* spectroscopic map (term as *I*(*t*, **r**)) where *I-t* curves are collected at each spatial spots within a selected region. We extract dominated rotating rate $\omega$ by counting the current jump steps of *I-t* spectra, which quantizes the rotational switching among the major quantum states (Figure S4). The rate map $\omega(\mathbf{r})$ derived from the *I-t* spectroscopic map *I*(*t*, **r**) illustrates that the rotating rates are notably high and primarily localized around the circular orbit region (left panel of Figure 1(e)).

We term the wheel-like protrusion in the STM image as a diatomic rotor, comprising simply two metal atoms, as depicted in Figure 1(f). The Cs atom positioned at the hexagonal center of Sb honeycomb lattice serves as an anchor, corresponding to the inner spherical center of wheel-like protrusion. The Cr atom rotates around the anchor along the axis within central Cs atom, forming a circular orbit region of the wheel-like protrusion.

To investigate the dynamics of the Cs-Cr diatomic rotor, we conduct the bias-dependent STM measurements of the Cr-Cs rotor. The topographic features undergo significant changes with the increasing sample bias (Figure 2(a) – (c)). At a low sample bias of -0.05 V, Cr-Cs rotors exhibit a 12-petals flower-shaped feature in the STM image (Figure 2(a)), suggesting 12 possible adsorption sites (Figure S5). With an increment in the sample bias, the topography of rotor gradually transforms into the wheel-like protrusion where annulus orbit become more unstable than those in low-bias STM images (Figure 2(b), (c)). It indicates that the switching rate of the diatomic rotor becomes larger with the increasing sample bias, which is further supported by the bias-dependent *I-t* spectra (Figure 2(d)). When applying the low sample bias of -0.05 V, the *I-t* curve acquired at the annulus orbit show no steps at the time scale of 250 ms, indicating a low switching rate. As the sample bias value increases (*e.g.* -0.2V and -0.6 V), the switching rates rise remarkably (Figure 2(d)). To reveal the evolution of the rate with the sample bias, we collect the *I-t* spectroscopic maps *I*(*t*, **r**) on a diatomic Cr-Cs rotor at various scanning bias. At each sample bias, we carry out a statistical analysis of the rotating rate $\omega$ extracted from the *I*(*t*, **r**) of rotor and get the spatial-averaged values $\bar{\omega}$, respectively. It clearly shows that the bias-dependent

$\bar{\omega}(V)$ exhibits a gap like feature where the rotating rate drops dramatically below a critical scanning bias ~0.2 V. Such gap feature demonstrates an activation energy of the rotation switching of the Cr-Cs rotor from STM tip (Figure 2(e)). In addition, the bias-dependent behavior is not symmetric for the positive and negative sample bias, indicating the polar boosting effect by the electric field (Figure S6).

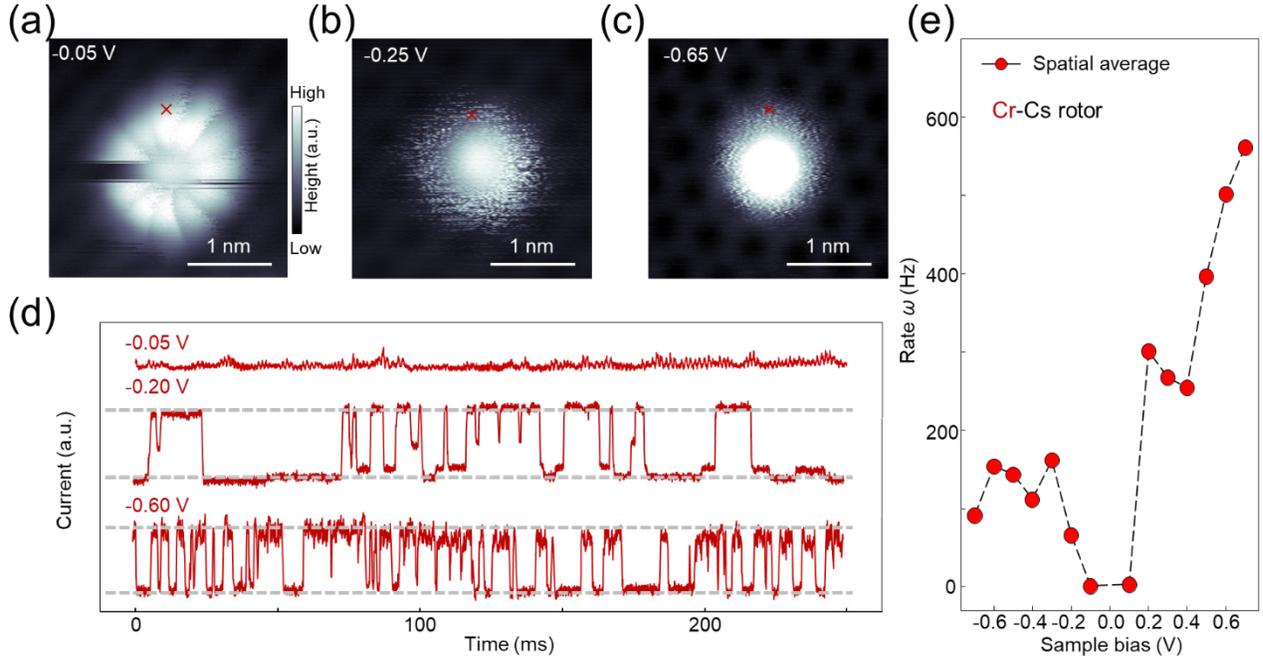

**Figure 2.** Bias-dependent rotational frequency of Cr-Cs rotors. The STM topography of a Cr-Cs rotor at sample bias of (a) -0.05 V, (b) -0.25 V and (c) -0.65 V, respectively, showing that the rotation switching of Cr atom gradually increases with larger sample bias ($I$=0.3 nA). (d), The *I-t* spectra measured at the annulus position (red cross) of Cr-Cs rotor in (a) - (c). In the same period, the current steps change slow at small bias but jump fast at large bias ($I$=0.1 nA). (e), The bias-dependent rotation switching rate $\bar{\omega}(V)$ of Cr-Cs rotor, showing a bias gap where the rotor freezes inside the gap. The rotation switching rate $\bar{\omega}(V)$ is spatially averaged over the rotor.

The formation of diatomic rotor is not limited by the Cr and Cs atoms but can be extended to other transition metal atoms and alkali metal atoms in the $AV_3Sb_5$ system. For instance, we have deposited the Fe atoms onto Sb surface of $CsV_3Sb_5$ with diluted Cs adatoms. The Fe adatoms capture the isolated Cs atoms on the Sb surface, giving rise to the formation of Fe-Cs diatomic rotors. The Fe-Cs rotors exhibit analogous features to the Cr-Cs rotors, identified as wheel-like protrusions in the STM image (Figure 3(a)). The rotating rate map $\omega(r)$, extracted from FT of $I$-$t$ map, illustrates six sites with large switching rates in the annulus orbit region of Fe-Cs rotor (Figure 3(b)), namely six most unstable sites for Fe atoms. The $I$-$t$ spectra at the annulus orbit of Fe-Cs rotor shows a step-like feature akin to the Cr-Cs rotor (Figure 3(c)). In addition, the bias-dependent mean rotating rate of Fe-Cs rotor $\bar{\omega}(V)$ also exhibits a bias gap feature, indicating the activation energy of the rotation of Fe-Cs rotor from the STM tip (Figure 3(d) and Figure S7).

It is noticeable that the unidirectional $4a_0$ charge stripe [31-36] (Figure S8(a)) in the Sb surface of $CsV_3Sb_5$ can effectively tuning the distribution of switching rates of the rotors. The radially switching rates distributed anisotropic (Figure S8(b)) respective to the $4a_0$ axis, suggesting the effects of symmetry breaking charge orders on rotors.

There are several notable distinctions between Fe-Cs and Cr-Cs rotors. Firstly, the range of current switching range is between -0.3 nA and -0.05 nA (Figure 3(c)), larger than that of Cr-Cs rotor (switching between -0.3 nA and -0.1 nA). Secondly, the bias gap size (~0.2 V) of Fe-Cs rotor is larger than that (~0.1 V) of Cr-Cs rotor. Finally, the $\bar{\omega}(V)$ is more symmetric for positive and negative sample bias of Fe-Cs rotor (Figure S9), in contrast to the asymmetric feature observed in the Cr-Cs rotor. These distinctions might result from variations in adatom absorption and distinct bond strengths between adatoms and Cs atoms. Beyond the magnetic elements, we also deposit the non-magnetic V atoms onto the Sb surface of $CsV_3Sb_5$ and $KV_3Sb_5$ to construct V-Cs (Figure 3(e)) and V-K (Figure 3(f)) diatomic rotor, respectively. The universal formation of transition metal-alkali metal diatomic rotors provides a versatile platform to modulate rotating behaviors by selecting specific elements.

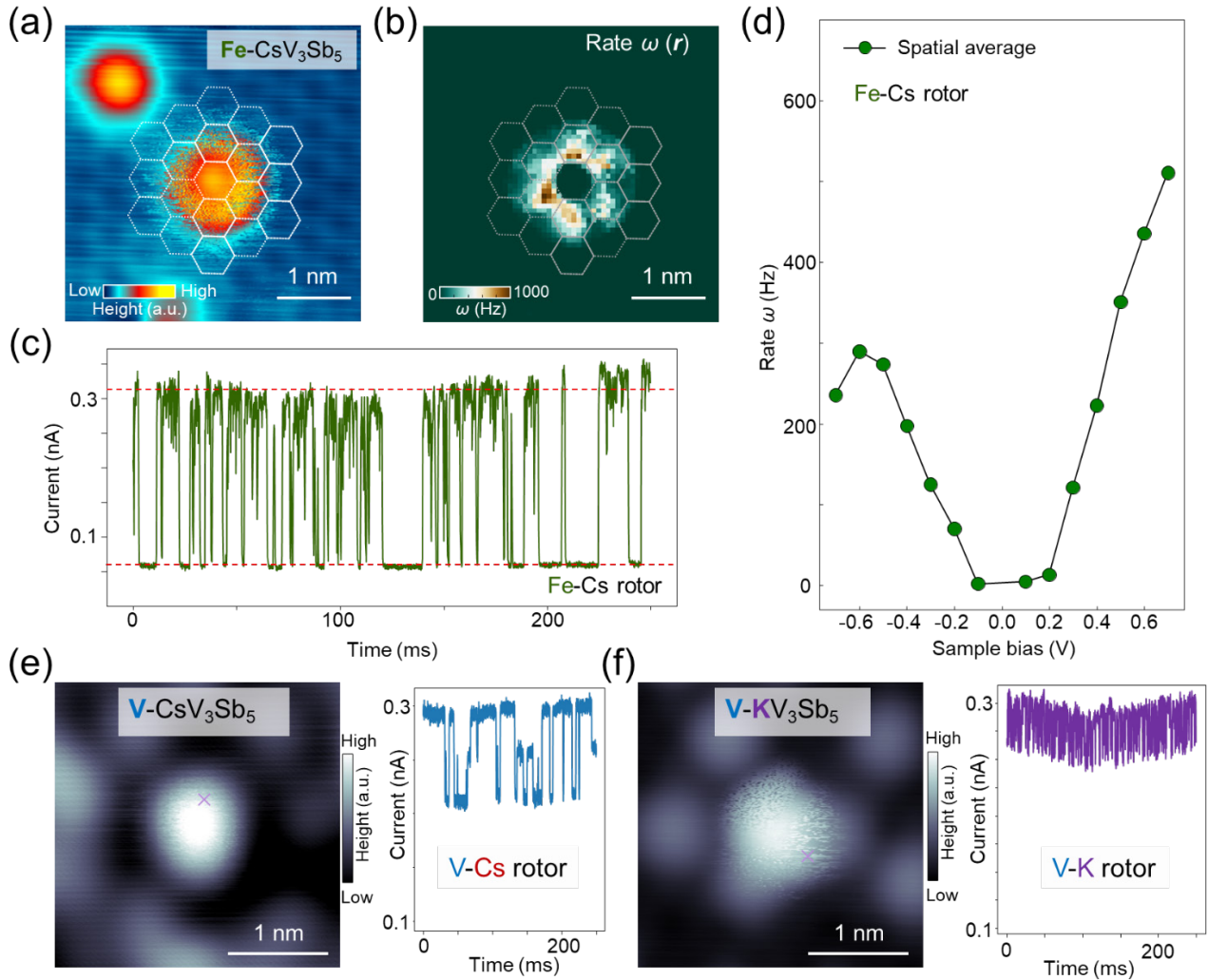

**Figure 3.** The formation of diatomic rotors in metal-$A$V$_3$Sb$_5$ ($A$= K, Cs) system. (a), The STM topography of Fe-Cs rotors in CsV$_3$Sb$_5$. ($V$=-400 mV, $I$=100 pA). (b), Frequency map of (a), showing that Fe atom rotates around Cs in six most unstable sites ($V$=-400 mV, $I$=100 pA). (b), The $I$-$t$ spectrum measured at the annulus position (red cross) of Fe-Cs rotor. ($V$=-600 mV, $I$=0.3 nA). (d), The bias-dependent frequency evolution of Fe-Cs rotor, showing a bias gap feature where the rotor freeze at small bias. The rotation switching rate $\bar{\omega}(V)$ is spatially averaged over the rotor. (e), The STM topography (left) and $I$-$t$ spectrum (right) of V-Cs rotors in CsV$_3$Sb$_5$. (Scan: $V$=-1 V, $I$=20 pA; $I$-$t$: $V$=-300 mV, $I$=300 pA). (f), The STM topography (left) and $I$-$t$ spectrum (right) of V-K rotors in KV$_3$Sb$_5$. (Scan: $V$=-200 mV, $I$=50 pA; $I$-$t$: $V$=-300 mV, $I$=300 pA).

The atomically control over the rotational switching of diatomic rotors is achieved by directly manipulating and assembling them by an STM tip. The Cs atoms, serving as the anchor of diatomic rotor, are easily moved by the STM tip at low temperature [35, 41]. Consequently, the local position of a rotor at the Sb surface is atomically manipulated as well. During the STM manipulation process, we initially approach the STM tip closer to the surface and position it at one side of the rotor. The tip is then moved across the rotor and towards a selected spatial position. Maintaining the position of one rotor unchanged, we move another rotor closer, creating an artificial system for studying the rotating behaviors of two adjacent rotors. Initially, the distance between two rotors is $3a_0$, leaving the rotation rate of each rotor undisturbed (Figure 4(a)). As the two rotors approach each other closely ($2a_0$), the central region hinders the rotary motion of the Fe atoms (Figure 4(b)), resulting in a stable *I-t* spectrum (Figure 4(d)). Moving even closer to a distance of $a_0$, the Fe atoms rotate around the two adjacent Cs atoms at much slower rates (Figure 4(c)) than a single rotor. Beyond the manipulation of two rotors, we achieve the assembly of the rotors into different motifs to modulate the rotating rates. For example, we assemble three rotors into a triangle pattern (Figure 4(e) – (g)) through step-by-step STM manipulations. When the distance between the rotors is far ($\geq 3a_0$), the distribution of the rotation rates remains weakly affected (Figure 4(e), (f)). Moving closer to $2a_0$, Fe atoms are restricted from the center of triangular motif, resulting in a triangle orbit of rotation (Figure 4(g)). The weak van der Waals interaction between the rotors and substrates provides the opportunity to artificially design and build assembled rotational nanostructures atom by atom [42, 43].

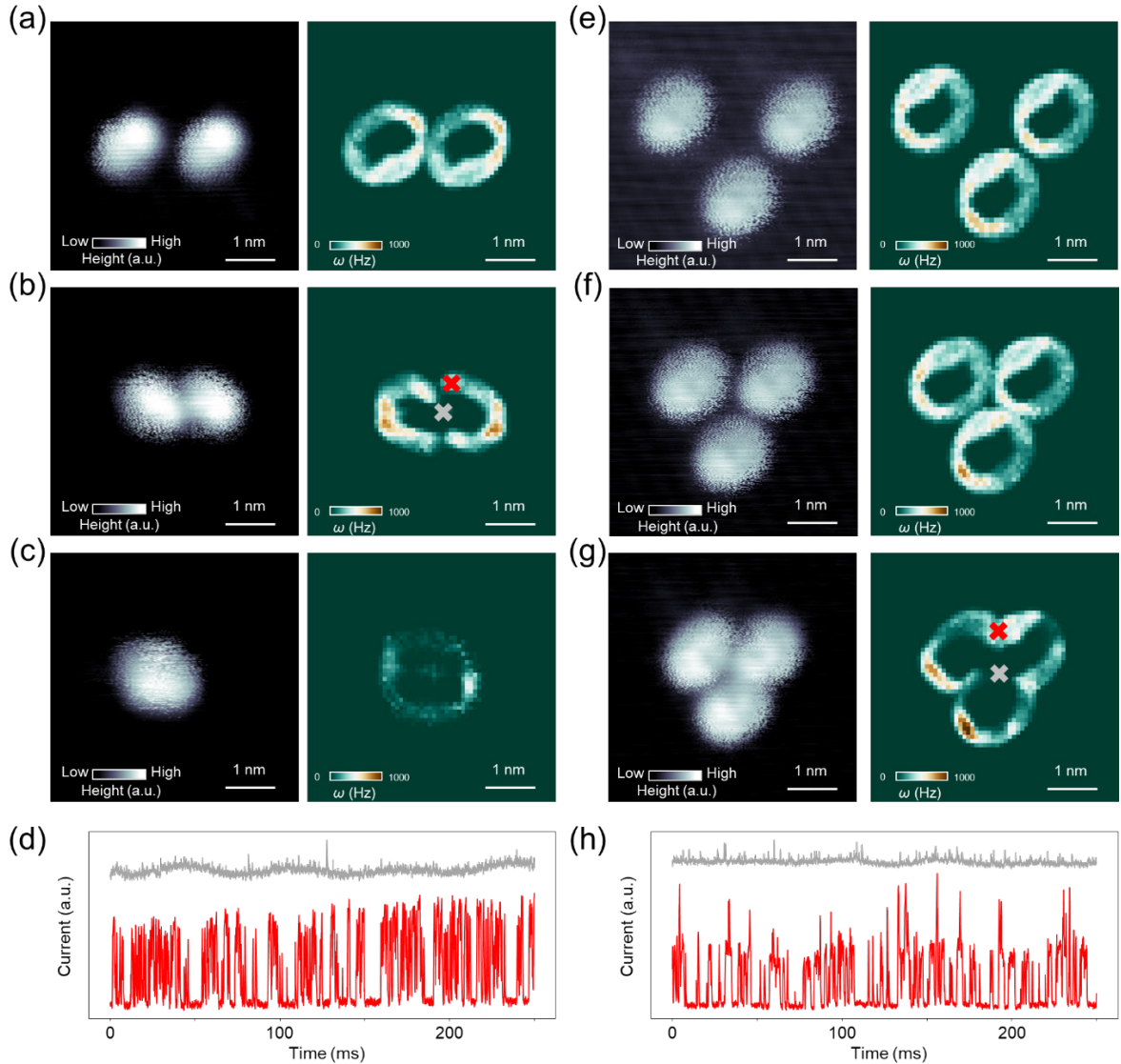

**Figure 4.** Atomically-engineering of Fe-Cs rotors by STM manipulation. (a), The STM topography (left panel) of two Fe-Cs rotors and their rates map (right panel). The distance between two rotors is $3a_0$ ($V=-400$ mV, $I=100$ pA). (b), The STM topography (left panel) of two Fe-Cs rotors and their rates map (right panel). The distance between two rotors is $2a_0$ ($V=-400$ mV, $I=100$ pA). (c), The STM topography (left panel) of two Fe-Cs rotors and their rates map (right panel). The distance between two rotors is $a_0$ ($V=-400$ mV, $I=100$ pA). (d), The $I$-$t$ spectra at the center (gray cross) and annulus (red cross) in (b), respectively, showing significant difference in rotating frequency. (e), The STM topography (left panel) of three Fe-Cs rotors and their rates map (right panel). The distance between the rotors is $5a_0$ and $4a_0$, respectively ($V=-400$ mV, $I=100$ pA). (f), The STM topography (left panel) of three Fe-Cs rotors and their rates map (right panel). The distance between rotors is $3a_0$ ($V=-400$ mV, $I=100$ pA). (g), The STM topography (left panel) of three Fe-Cs rotors and their rates map (right panel). The distance between rotors is $2a_0$ ($V=-400$ mV, $I=100$ pA). (h), The $I$-$t$ spectra at the center (gray cross) and annulus (red cross) in (g), respectively, showing significant difference in rotating frequency.

The realization of diatomic rotors introduces a novel paradigm for the design and construction of building blocks crucial for functional nanomachines and the future landscape of nanotechnology. The inherent simplicity of diatomic rotors marks the inception of atomic machines, serving as fundamental components for more intricate nano-machinery. The diatomic rotors consist of magnetic elements implicate the applications in the nanoscale robots based on spintronics. Furthermore, the electrical controllability of the on/off rate in diatomic rotors opens avenues for potential electrical applications. The symmetry-breaking rotation switching of the rotor probably correlate to the symmetry breaking charge orders [34] in the kagome superconductors, which provides possibilities to tune the electronic states of the rotors by the degrees of charges. Substituting atoms within the diatomic rotors allows for the tuning of their quantum properties, establishing a novel platform for exploring two-body quantum interactions. The discovery is poised to propel advancements in the design of functional and automated nanomachinery, particularly those capable of operation on material surfaces [8]. The potential for tuning quantum properties and leveraging electronic control underscores the versatility of diatomic rotors as foundational elements for emerging nanotechnological applications.

# EXPERIMENTAL METHOD

**Sample preparation.** Single crystals of $CsV_3Sb_5$ and $KV_3Sb_5$ were grown *via* a modified self-flux method [35]. The samples used in the STM/S experiments are cleaved in situ at low temperature (13 K) and immediately transferred to an STM chamber. Cr, Fe and V were *in situ* evaporated from pure metal rods (purity of 99.99%) using a commercial electron beam evaporator and deposited onto the surface of the $CsV_3Sb_5$ with a substrate temperature of ~20 K.

**Scanning tunneling microscopy/spectroscopy.** Experiments were performed in an ultrahigh vacuum ($1\times10^{-10}$ mbar) ultra-low temperature STM system equipped with 11 T magnetic field. Non-superconducting tungsten tips were fabricated via electrochemical etching and calibrated on a clean Au(111) surface prepared by repeated cycles of sputtering with argon ions and annealing at 500 °C. STM topographic images are acquired in constant-current mode, with all the scanning parameters (setpoint voltage $V_s$ and current $I_t$) listed in the figure captions. Unless otherwise noted, the d$I$/d$V$ spectra were acquired in constant-height mode by a standard lock-in amplifier at a modulation frequency of 973.1 Hz. Current−time ($I$−$t$) curves, for measuring the occupation probability of adjacent rotational configurations of the diatomic rotors, were acquired at a selected spatial spot with the feedback loop closed. The *I-t* map $I(\mathbf{r}, t)$ were acquired in constant-height mode where the *I-t* curves were collected at each spatial spots of the map. The frequency map $f(\mathbf{r})$ was obtained by extracting the frequency value $f$ from the *I-t* curves in each spatial spot of the *I-t* map by counting the step changing (details are shown in Figure S4).

# AUTHOR INFORMATION

**Corresponding Authors**

**Hui Chen** hchenn04@iphy.ac.cn,

**Hong-Jun Gao** hjgao@iphy.ac.cn